\documentclass[bibnotes,twocolumn, showpacs,aps,prb,a4paper]{revtex4}

	\usepackage{color,epsfig,rotating,graphicx,subfigure}
\usepackage{amsmath,amssymb,amscd}

\usepackage[latin1]{inputenc}
\usepackage[english]{babel}

\usepackage{dsfont}
\usepackage{textcomp}

\renewcommand{\Re}{{\rm Re}}
\renewcommand{\Im}{{\rm Im}}

\newcommand{\ri}{{\rm i}}
\newcommand{\re}{{\rm e}}
\newcommand{\rd}{{\rm d}}
\newcommand{\rr}{{\rm r}}

\newcommand{\rl}{{\rm l}}

\newcommand{\rs}{{\rm s}}
\newcommand{\rp}{{\rm p}}  

\newcommand{\kb}{k_{\rm B}}

\newcommand{\Tr}{{\rm Tr}}

\newcommand{\rth}{{\rm th}}
\newcommand{\coh}{{\rm coh}}

\begin{document}

%
%
\title{Influence of roughness on the near-field heat transfer between two plates}

\author{S.-A. Biehs and J.-J. Greffet}

\affiliation{Laboratoire Charles Fabry, Institut d'Optique, CNRS, Universit\'{e} Paris-Sud, Campus
Polytechnique, RD128, 91127 Palaiseau cedex, France}

\date{16.11.2010}
\begin{abstract}
The surface roughness correction to the near-field heat transfer between two rough bulk materials is discussed
by using second-order perturbation theory. The results allow for estimating the impact of surface 
roughness to the heat transfer in recent experiments between two plates and between a micro-sphere and a plate (using the
Derjaguin approximation). Furthermore, we show that the proximity approximation for describing rough surfaces 
is valid for distances much smaller than the correlation length of the surface roughness 
even if the heat transfer is dominated by the coupling of surface modes. 
\end{abstract}

\maketitle

\newpage

%
%

\section{Introduction}

Recently, several experimental setups made a measurement of the radiative heat flux
on nanoscale feasible~\cite{HuEtAl2008,NarayaEtAl2008,ShenEtAl2008,RousseauEtAl2009}. 
It could be verified by Hu {\itshape et al.} that the heat flux between two glass plates exceeds the far-field limit set by 
Planck's black body radiation law~\cite{HuEtAl2008} for distances of some microns. 
Narayanaswamy {\itshape et al.} and Shen {\itshape et al.} used a different experimental  
setup measuring the heat flux between microspheres and plates of different materials which allow 
for detecting the radiative heat flux at much smaller distances and have 
reported heat transfer coefficients three orders of magnitude larger than the black body 
radiation limit~\cite{NarayaEtAl2008,ShenEtAl2008} in accordance with theoretical predictions~\cite{JoulainEtAl2005,VolokitinPersson2007}. 
A similar setup was used by Rousseau {\itshape et al.} for measuring the 
heat flux between a glass microsphere and a glass sample in a distance regime ranging from
$30\,{\rm nm}$ to $2.5\mu{\rm m}$~\cite{RousseauEtAl2009}. With that experiment the theoretical 
predictions based on fluctational electrodynamics~\cite{PvH1971} could be verified with high accuracy.  

In all these experiments the results have been compared to calculations which do not take into account the
roughness of the materials used. 
Here, we want to tackle the question how the surface roughness affects the
heat flux in the near-field. Therewith we provide the basis for comparison of the experimental data with 
theoretical results including roughness effects. Furthermore, our results might be used to study the impact of
roughness in thermophotovoltaic devices~\cite{MatteoEtAl2001, NarayanaswamyChen2003, LarocheEtAl2006, FrancoeurEtAl2008, ZhangReview}.
The first work considering surface roughness effects for the near-field heat flux was 
given by Persson {\itshape et al.}~\cite{PerssonEtAl2010} employing the so-called proximity approximation (PA)~\cite{Derjaguin1934, BlockiEtAl1977} to determine the
heat flux between two rough surfaces. In a recent work~\cite{BiehsGreffet2010} considering the effect of roughness on the heat flux between a nanoparticle
and a rough surface, it has been shown that perturbation
theory exactly reproduces the PA for distances smaller than the correlation length of the rough surface. 

In this work, we extend the perturbation theory to describe the near-field heat transfer between two
semi-infinite media with rough surfaces. This geometry is more suitable for the estimation of surface roughness
effects in recent experimental setups~\cite{HuEtAl2008}. Utilizing the Derjaguin approximation, our results can also 
be used to calculate the roughness correction for the experiments of Narayanaswamy {\itshape et al.}, Shen {\itshape et al.}~\cite{NarayaEtAl2008,ShenEtAl2008} 
and Rousseau {\itshape et al.}~\cite{RousseauEtAl2009}. 
In addition, we show that the PA can be used for vacuum gaps smaller than the correlation length of the surface roughness even 
when the heat flux is dominantly due to the coupling of surface modes. 
This is an important result since it shows that the PA can indeed be used to obtain a simple estimation of the roughness correction to the heat flux. 

The paper is organized as follows: In Sec.\ II we introduce the mean Poynting vector in terms of transmission coefficients and present 
in Sec.\ III the resulting expressions for second-order perturbation theory considering a Gaussian surface roughness. Finally, in Sec.\ IV 
we study the impact of surface roughness on the heat flux and give a detailed discussion of the numerical results. In particular, we investigate
the range of validity of the PA.

%
%

\section{Radiative heat transfer}

\begin{figure}[Hhbt]
  \epsfig{file=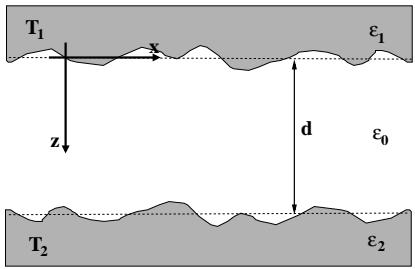, width=0.3\textwidth}
  \caption{\label{Fig:RoughRough} Sketch of the situation considered here.}
\end{figure}

Let us consider a configuration as depicted in Fig.~\ref{Fig:RoughRough}. We have two semi-infinite
bodies in local thermal equilibrium at temperatures $T_1$ and $T_2$ separated by a vacuum
gap. In general both media have different material properties which can be expressed by different
permittivies $\epsilon_1$ and $\epsilon_2$. When considering isotropic and local materials only, the
permittivities are scalars and do not depend on the wave vector. Here, we will consider
such materials, although in general non-local effects have to be taken into account~\cite{ChapuisEtAl2008,JoulainHenkel2005}. 

Furthermore, we assume that both bodies have a rough surface expressed by the surface
profile functions $S_1(\mathbf{x})$ and $S_2(\mathbf{x} + d)$, where $d$ is the mean distance
between the two materials and $\mathbf{x} = (x,y)$. Since in this work we focus on stochastic Gaussian surface 
profiles~\cite{Beckmann1963}, the profile functions fullfill the properties  
\begin{align}
  \langle S_i (\mathbf{x}) \rangle &= 0, \\
  \langle S_i (\mathbf{x}) S_i(\mathbf{x}') \rangle &= \delta^2_i W_i (|\mathbf{x - x'}|).
\end{align}
The brackets $\langle \rangle$ stand for the average over an ensemble of realizations of
the surface profiles $S_i(\mathbf{x})$ for $i=1,2$; $\delta_i$ is the rms height
of the surface profiles. The correlation functions $W_i(|\mathbf{x - x'}|)$ are given
by a Gaussian
\begin{equation}
  W_i (|\mathbf{x - x'}|) = \re^{-\frac{|\mathbf{x - x'}|^2}{a^2_i}}
\end{equation}
introducing the transverse correlation length $a_i$. In addition we assume that both surface profiles 
are statistically independent, i.e., $\langle S_1 S_2 \rangle = 0$.
For the Fourier component $\tilde{S}_i(\boldsymbol{\kappa})$ of the surface profile functions one obtains for $i = 1,2$
\begin{align}
  \langle \tilde{S}_i (\boldsymbol{\kappa}) \rangle &= 0 \\
  \langle \tilde{S}_i (\boldsymbol{\kappa}) \tilde{S}_i (\boldsymbol{\kappa}')\rangle &= (2 \pi)^2 \delta^2_i 
                      \delta(\boldsymbol{\kappa} + \boldsymbol{\kappa}') g_i (\kappa)
\end{align}
with the surface roughness power spectra
\begin{equation}
  g_i (\kappa) = \int\!\!\rd^2 x\, W_i (|\mathbf{x}|) \re^{- \ri \boldsymbol{\kappa} \cdot \mathbf{x}} = \pi a^2_i \re^{-\frac{\kappa^2 a^2_i}{4}}.
\label{Eq:RoughnessPowerSpec}
\end{equation}

In order to evaluate the radiative heat transfer between the two bodies~\cite{VolokitinPersson2007}, it suffices to 
determine the electric Green's dyadic $\mathds{G}(\mathbf{r,r'})$ with $\mathbf{r}$ and
$\mathbf{r}'$ within the vacuum gap. Then the mean Poynting vector is given by~\cite{VolokitinPersson2007}
\begin{equation}
  \langle S \rangle = \int\!\!\frac{\rd \omega}{2 \pi} \biggl[ \Theta(\omega,T_1) - \Theta(\omega,T_2) \biggr] \langle  S_\omega \rangle
\label{Eq:PoyntRough_a}
\end{equation}
with
\begin{equation}
\begin{split}
 \langle  S_\omega \rangle &= 2 \Re\Tr \biggl[ \int \!\!\rd^2 x' \biggl\langle \mathds{G}(\mathbf{r,r'}) \partial_z \partial_{z'} {\mathds{G}}^\dagger(\mathbf{r,r'}) \\
                               &\quad - \partial_z {\mathds{G}}^\dagger(\mathbf{r,r'}) \partial_{z'} \mathds{G}(\mathbf{r,r'}) \biggr\rangle \biggr]_{\mathbf{r} = \mathbf{r}'}
\end{split}
\label{Eq:PoyntRough_b}
\end{equation}
where the integration is carried out over a flat surface within the vacuum gap parallel to the $x$-$y$ plane
and
\begin{equation}
  \Theta(\omega,T_i) = \frac{\hbar \omega}{\re^{\hbar \omega \beta_i} - 1}
\end{equation}
with $\beta_i = 1/(\kb T_i)$ for $i = 1,2$; $\kb$ is Boltzmann's constant.
Finally, we reformulate Eq.~(\ref{Eq:PoyntRough_b}) in terms of transmission coefficients. To this end, we first introduce 
the Fourier representation of the Green's dyadic as 
\begin{equation}
  \mathds{G}(\mathbf{r,r'}) = \int \!\!\frac{\rd^2 \kappa}{(2 \pi)^2} \int\!\!\frac{\rd^2 \kappa'}{(2 \pi)^2} \, \re^{\ri (\boldsymbol{\kappa} \cdot \mathbf{x} - \boldsymbol{\kappa}'\cdot\mathbf{x}')} \mathds{G} (\boldsymbol{\kappa},\boldsymbol{\kappa}';z,z')
\end{equation}
with $\mathbf{x} = (x,y)$ and the lateral wave vector $\boldsymbol{\kappa} = (k_x, k_y)$.
Then we find for $\langle S_\omega \rangle$ in Eq.~(\ref{Eq:PoyntRough_b})
\begin{equation}
  \langle S_\omega \rangle = \sum_{j,j' = {s,p}} \int\!\!\frac{\rd^2 \kappa}{(2 \pi)^2} \int\!\!\frac{\rd^2 \kappa'}{(2 \pi)^2} \re^{\ri (\boldsymbol{\kappa} - \boldsymbol{\kappa}') \cdot \mathbf{x}} \langle T_{j,j'} (\boldsymbol{\kappa},\boldsymbol{\kappa}'; \omega) \rangle,
\label{Eq:PoyntRough_b_transmission}
\end{equation}
where
\begin{equation}
\begin{split}
   T_{j,j'} (\boldsymbol{\kappa},\boldsymbol{\kappa}'; \omega) &= 2 \Re\Tr \!\! \int \!\! \frac{\rd^2 \kappa''}{(2 \pi)^2} 
                             \biggl[ \mathds{G}(\boldsymbol{\kappa},\boldsymbol{\kappa}'') \partial_z \partial_{z'} {\mathds{G}}^\dagger(\boldsymbol{\kappa}',\boldsymbol{\kappa}'') \\
                             &\quad - \partial_z {\mathds{G}}^\dagger(\boldsymbol{\kappa},\boldsymbol{\kappa}'') \partial_{z'} \mathds{G}(\boldsymbol{\kappa}',\boldsymbol{\kappa}'') \biggr]_{z = z'} 
\end{split}
\label{Eq:TransmissionCoefficient}
\end{equation}
is the transmission coefficient. It describes the transmission of a thermally emitted wave with wave vector $\boldsymbol{\kappa}$ 
and polarization $j$ from body $1$ to body $2$ 
which is scattered into wave vector $\boldsymbol{\kappa}'$ and polarization $j'$ during the transmission process.   
By means of the transmission coefficient the
Poynting vector in Eq.~(\ref{Eq:PoyntRough_a}) can be converted into a Landauer-like form as was shown in Ref.~\cite{GreffetEtAl2010}.
After ensemble averaging, the translational and rotational symmetries are restored
so that the mean transmission coefficient reduces to
\begin{equation}
  \langle T_{j,j'} (\boldsymbol{\kappa},\boldsymbol{\kappa}'; \omega) \rangle = (2 \pi)^2 \delta(\boldsymbol{\kappa} - \boldsymbol{\kappa}') \delta_{j,j'} T_j (\boldsymbol{\kappa};\omega).
\label{Eq:MeanTransmissionCoefficient}
\end{equation}
and hence
\begin{equation}
 \langle S_\omega \rangle =  \sum_{j = {s,p}} \int\!\!\frac{\rd^2 \kappa}{(2 \pi)^2} T_{j} (\boldsymbol{\kappa}; \omega).
\label{Eq:PoyntSpecTrans}
\end{equation}
In the following section we will determine the perturbation expansion of the transmission coefficient $T_j = T_j^{(0)} + T_j^{(1)} +T_j^{(2)} + \ldots$ and $\langle S_\omega \rangle =  S_\omega^{(0)} + \langle S_\omega^{(1)} \rangle + \langle S_\omega^{(2)} \rangle + \ldots$

\subsection{Perturbation result}

By using Eqs.~(\ref{Eq:TransmissionCoefficient})-(\ref{Eq:PoyntSpecTrans}) we obtain the perturbation expansion for the mean Poynting vector by
expanding the Green's dyadic with respect to the surface profile functions $S_1$ and $S_2$. A short discussion of the perturbation method~\cite{Greffet1988} used here,
can be found in Ref.~\cite{BiehsGreffetsuppl2010}. Note, that this perturbation approach has also been used to study for example the roughness effects for 
the Casimir force~\cite{MaiaNetoEtAl2005,LamrechtEtAl2006}. 
We remark that the perturbation theory can be applied as far as the rms of the surface profiles
is the smallest lenght scale of the problem~\cite{HenkelSandoghdar1998}. 

Then, by inserting the zeroth-order dyadic Green's function from~\cite{BiehsGreffetsuppl2010} into the expression for
the mean transmission coefficient in Eq.~(\ref{Eq:TransmissionCoefficient}) one retrieves for $\langle S_\omega \rangle$ from 
Eq.~(\ref{Eq:PoyntSpecTrans}) 
the well-known result~\cite{PvH1971} for the case of two media with flat surfaces
\begin{equation}
  {T_j^{(0)}} (\kappa;\omega)  = 
                            \begin{cases}  
                               \frac{(1 - |r_j^1|^2)(1 - |r_j^2|^2)}{|D_{jj}|^2}             &, \kappa \leq k_0 \\
                               \frac{4 \Im(r_j^1)\Im(r_j^2) \re^{- 2 \gamma d}}{|D_{jj}|^2} &, \kappa > k_0 
                            \end{cases} ,
\label{Eq:DefI0b}
\end{equation}
where $\gamma^2 = \kappa^2 - k_0^2$,  $k_0 = \omega/c$ and $r_{j}^i$ ( for $i = 1,2$ and $j = \rs,\rp$) are Fresnel's reflection coefficients
of the two bodies for $\rs$ or $\rp$ polarization;  $D_{jj} = 1 - r_j^1 r_j^2 \exp(2 \ri \gamma_\rr d)$  is the Fabry-P\'{e}rot like
denominator where $\gamma_\rr^2 = k_0^2 - \kappa^2$. 
$T^{(0)}_j$ is the transmission coefficient between the two bodies with flat surfaces having
values between zero and one. In the propagating regime ($\kappa < k_0$) the property $T^{(0)}_j \leq 1$ means that the
heat flux is always smaller than that between two black bodies for which $T^{(0)}_j = 1$. 
In the evanescent regime the maximal value of one can be achieved for such pairs $(\omega,\kappa)$
which fullfill the dispersion relation of the coupled surface modes of the two surfaces~\cite{Pendry1999}.

To get the first nonvanishing term for the surface roughness correction, it is necessary to expand the Green's functions up to second order,
since $\langle \mathds{G}^{(1)} \rangle = 0$ and therefore ${T^{(1)}_j} = \langle S_\omega^{(1)} \rangle = 0$. The resulting
second-order correction can be written as
\begin{equation}
\begin{split}
    \langle S_\omega^{(2)} \rangle &= S^{(2)}_{\omega,\rm diff} + S^{(2)}_{\omega,\rm spec}  \\
                                       &= \sum_{j = \rs,\rp} \int\!\!\frac{\rd^2 \kappa}{(2 \pi)^2} \, ({T^{(2)}_{j,{\rm diff}}} + {T^{(2)}_{j,{\rm spec}}}). 
\end{split}
\label{Eq:SpecSecondOrder}
\end{equation}
where we have splitted the result into the specular part which depends on the mean field or mean Green's function only, and the so-called
diffuse part which is per definition given by $\langle S_\omega^{(2)} \rangle - S^{(2)}_{\omega,\rm spec}$.
For the diffuse part, we obtain~\cite{BiehsGreffetsuppl2010}
\begin{equation}
\begin{split}
    \sum_{j = \rs,\rp} {T^{(2)}_{j,{\rm diff}}}  
                            &= - \int \!\! \frac{\rd^2 \kappa'}{(2 \pi)^2} 
                                 k_0^2 |\epsilon_1 - 1|^2 \frac{g_1(|\boldsymbol{\kappa} - \boldsymbol{\kappa}'|) (k_0 \delta_1)^2}{4 |\gamma_\rr|^2 |\gamma_\rr'|^2}
                                \\
                            &\qquad\times\biggl(  Q_\rs^2 {Q_\rs^2}' (\hat{\boldsymbol{\kappa}}\cdot\hat{\boldsymbol{\kappa}}')^2 
                              + Q_\rs^2 {Q_\rp^2}' \frac{|\gamma_1'|^2}{|\epsilon_1| k_0^2} (\hat{\boldsymbol{\kappa}}\times\hat{\boldsymbol{\kappa}}')^2 \\
                            &\qquad\qquad   + Q_\rp^2 {Q_\rs^2}' (\hat{\boldsymbol{\kappa}}\times\hat{\boldsymbol{\kappa}}')^2 \frac{|\gamma_1|^2}{|\epsilon_1| k_0^2} \\
                            &\qquad\qquad   + Q_\rp^2 {Q_\rp^2}' \frac{|\kappa\kappa' \epsilon_1 - \hat{\boldsymbol{\kappa}}\cdot\hat{\boldsymbol{\kappa}}' \gamma_1 \gamma_1'|^2}{|\epsilon_1|^2 k_0^4} \biggr) \\
                            &\quad + (1 \leftrightarrow 2),
\end{split}
\label{Eq:PoyntingDiff}
\end{equation}
where $t_{\rs/\rp}^i$ for $i = 1,2$ are the amplitude transmission coefficients of the two interfaces 
for s- and p-polarized modes; $\gamma_i^2 = k_0^2 \epsilon_i - \kappa^2$ ($i = 1,2$), 
$\hat{\boldsymbol{\kappa}} = (k_x,k_y)/\kappa$ and $(1 \leftrightarrow 2)$ symbolizes the term obtained by interchanging
the index $1$ and $2$. 
The functions $Q_{\rs/\rp}^{1/2}$ are defined as
\begin{align}
Q_{\rs/\rp}^{1/2} &= \frac{|t_{\rs/\rp}^{2/1}|^2}{|D_{\rs\rs/\rp\rp}|^2} \biggl[ \Re(\gamma_\rr)\bigl(1 - |r_{\rp/\rs}^{1/2}|^2\bigr) \\
                  &\qquad\qquad\qquad     + 2 \Im(\gamma_\rr) \Im\bigl(r_{\rs/\rp}^{1/2}\bigr) \re^{-2 \gamma d} \biggr] \nonumber .
\end{align}
It is seen from Eq.~(\ref{Eq:PoyntingDiff}) that the diffuse correction to the transmission coefficient
and therefore also to $S^{(2)}_{\omega,\rm diff}$ is always negative. 
Hence, due to the diffuse scattering in the rough surface
the transmission coefficient becomes smaller and therefore the heat transfer less efficient.

For the specular contribution to the mean transmission coefficient we obtain~\cite{BiehsGreffetsuppl2010}  for propagating modes ($\kappa \leq k_0$) 
\begin{equation}
\begin{split}
  {T^{(2)}_{j,{\rm spec}}} &= -\frac{(1 - |r_j^2|^2)}{|D_{jj}|^2} 
                                   \Im \biggl[(r_j^1 - {r_j^2}^* \re^{-2\ri\gamma_\rr d})\frac{({t_j^1}^*)^2 {h_j^1}^* A_{1,jj}^*}{\gamma_\rr}
                                   \biggr]\\
                            &\quad+ (1 \leftrightarrow 2)
\end{split}  
\label{Eq:PertResSpecprop}
\end{equation}
and for evanescent modes ($\kappa > k_0$)
\begin{equation}
\begin{split}
  {T^{(2)}_{j,{\rm spec}}}
                            &= \frac{2 \Im(r_j^2)}{|D_{jj}|^2}  
                               \Im \biggl[(1 - r_j^1 {r_j^2}^* \re^{-2\gamma d})\frac{({t_j^1})^2 h_j^1 A_{1,jj}}{\gamma}\biggr] \re^{-2 \gamma d} \\
                            &\quad+ (1 \leftrightarrow 2).
\end{split}  
\label{Eq:PertResSpec}
\end{equation}
The expression for $A_{1/2,\rs\rs}$ and $A_{1/2,\rp\rp}$ can be found in 
Ref.~\cite{BiehsGreffetsuppl2010} and $h_\rs^{1/2} = 1, h_\rp^{1/2} = \epsilon_{1/2}$. 
By comparing the specular second-order transmission coefficents with the zeroth-order expressions, it
becomes apparent that they are very similar to ${T^{(0)}_j}$, but with the difference that one of the terms 
$1 - |r_j^{1/2}|$ or $2\Im(r_j^{1/2})$ is replaced by a much more complex term involving $A_{1/2,jj}$. By means of this term the
roughness scattering within one of the two surfaces 1 or 2 is taken into account yielding a roughness correction to the transmission which
can be either positive or negative. 

In general, within second-order perturbation theory the diffuse and the specular components 
describe the correction to the heat flux due to the scattering within only one of the rough surfaces, 
since both expressions split up into the sum of a term proportional to $\delta_1^2$ and $\delta_2^2$.
The different scattering processes can be classified as follows: The zeroth-order scattering is
the specular scattering of the two flat mean surfaces. 
The diffuse component describes the correction to the heat flux for first-order scattering of an incoming wave with lateral wave 
vector $\boldsymbol{\kappa}$ into a wave with $\boldsymbol{\kappa}'$. 
For such scattering processes, incoming s-polarized waves can be scattered into 
 p-polarized waves and propagating waves into evanescent
waves and vice versa. On the other hand, the specular component describes the correction to the heat flux due to a 
second-order scattering of an incoming wave with $\boldsymbol{\kappa}$
into a wave with $\boldsymbol{\kappa}'' = \boldsymbol{\kappa}$, since it describes the scattering of the mean field. 
These scattering processes can be divided into two different types: a direct second-order scattering 
originating from terms proportional to $\tilde{S}_i^{(2)}(\boldsymbol{\kappa} - \boldsymbol{\kappa}'')$  
and an indirect second-order scattering through intermediate
states with wave-vectors $\boldsymbol{\kappa}'$ originating from terms proportional to 
$\tilde{S}_i^{(1)} (\boldsymbol{\kappa} - \boldsymbol{\kappa}') \tilde{S}_i^{(1)} (\boldsymbol{\kappa}' - \boldsymbol{\kappa}'')$. The
latter one is a sequence of two first-order scattering processes as for the diffuse scattering, but with the constraint that the polarization 
and the lateral wave vector are the same before and after the scattering sequence.

%
%

\section{Numerical results and Discussion}

In the following we discuss the impact of surface roughness on the near-field heat transfer
numerically. We use the material properties of SiC and Au for both media with 
parameters taken from Refs.~\cite{ShchegrovEtAl2000,ChapuisEtAl2008} and set $T_2 = 0\,{\rm K}$ and $T_1 = 300\,{\rm K}$.
Additionally, we choose the surface roughness parameters $\delta_i = 5\,{\rm nm}$ and 
$a_i = 200\,{\rm nm}$ for $i = 1,2$, i.e.,
we consider a very shallow surface roughness with $\delta_i/a_i = 0.025$. 

\subsection{Distance dependence}

\begin{figure}[Hhbt]
  \epsfig{file=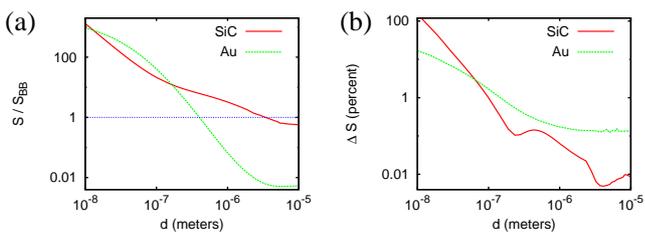, width=0.49\textwidth}
  \caption{\label{Fig:P0} (Color online)
     Plot of (a) $S^{(0)}$ over distance for two SiC plates and two Au plates normalized
     to the black body value $S_{\rm BB}$ with (b) the corresponding roughness correction $\Delta \langle S\rangle$. 
  }
\end{figure}

First, we turn to the distance dependence of the roughness correction to the near-field heat transfer. To this
aim, we start with a plot of the heat flux between two bodies with flat surfaces in Fig.~\ref{Fig:P0} (a)
considering two SiC and two Au slabs. SiC has a surface phonon resonance, whereas Au has no surface plasmon resonance in the infrared region. 
One can see that the curves for the SiC and the Au plates are quite different. For SiC 
the heat transfer is relatively large in the propagating regime ($d > \lambda_\rth = \hbar \beta c = 7.68\, \mu{\rm m}$ for $T = 300\,{\rm K}$) 
due to the low reflectivity of this material
and has values close to the black body limit $S_{\rm BB}= 459.3\, {\rm W}{\rm m}^{-2} $, whereas for Au which has a high reflectivity the heat flux is
very small. In the evanescent regime ($d < \lambda_\rth$) the heat flux is for
SiC dominated by the p-polarized surface mode contribution giving values which can exceed the black body result by
more than 3 orders of magnitude for the here chosen distance regime. On the other hand, for the Au plates the heat flux 
is dominated by the s-polarized contribution and the large amount of transfered heat can be attributed to induced Foucault currents~\cite{ChapuisEtAl2008}. 
For distances smaller than $100\,{\rm nm}$ 
the curve for SiC becomes proportional to $1/d^2$ which is the well-known result in the quasi-static limit~\cite{LevinEtAl1980,LoomisMaris1994}. 

Now, we have a look on the impact of surface roughness. In Fig.~\ref{Fig:P0} (b) we plot the roughness correction to the heat transfer
$\Delta S = \langle S^{(2)} \rangle /S^{(0)}$.
First of all, for both materials the roughness correction is positive for all distances so that roughness increases the
heat flux with respect to the case of flat surfaces. In the propagating regime the relative correction is very small (smaller than $0.01$ percent) for SiC and is limited due to the
fact that even if we introduce roughness the overall amount of thermal radiation cannot be larger than $S_{\rm BB}$. For Au the relative correction is one order of magnitude larger,
but the absolute correction is still very small, since $S^{(0)}$ has values smaller than $0.01 \cdot S_{\rm BB}$. In the opposite limit
of small distances, i.e., in the evanescent regime the relative correction due to roughness can become large. For example, for SiC the relative
correction is about $8$ percent for $d = 40\,{\rm nm}$. Since for that distance the heat flux is as large as $91$ times the black 
body value, this is an increase of about $7$ times $S_{\rm BB}$. For the Au plates 
we find a roughness correction of $5$ percent for $d = 40\,{\rm nm}$. At this distance $S^{(0)}$ is about $223$ times the black body value yielding 
an absolute correction of about $11$ times $S_{\rm BB}$.

\begin{figure}[Hhbt]
  \epsfig{file=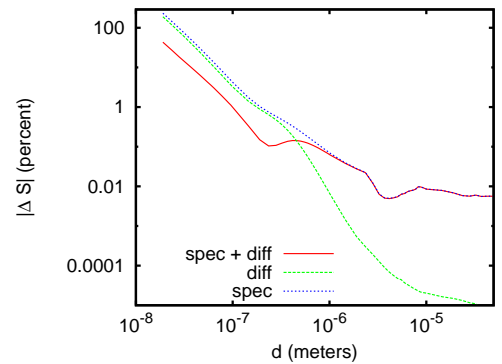, width=0.4\textwidth}
  \caption{\label{Fig:ProughSpecDiff} (Color online) 
     Plot of the specular and diffuse part of the roughness correction $|\Delta \langle S\rangle|$ 
     over distance for SiC.
   }
\end{figure}

Finally, in Fig.~\ref{Fig:ProughSpecDiff} we plot the specular and diffuse part of the roughness correction for SiC only.
Keeping in mind that the diffuse part is purely negative in this case, whereas we find that the specular part is purely positive. 
It can be seen that for distances larger than $500\,{\rm nm}$ the diffuse part is negligibly small compared to the specular one,
whereas for distances smaller than $500\,{\rm nm}$ both contributions have to be accounted for. Obviously, the specular
part tends to overestimate the roughness correction for such distances and only when also considering the diffuse contribution
one will get the correct result.

\begin{figure}[Hhbt]
  \epsfig{file=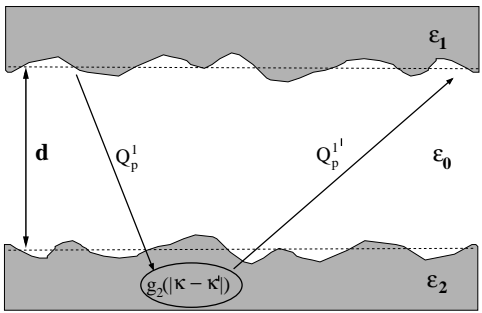, width=0.35\textwidth}
  \caption{\label{Fig:DiffusePart} Sketch of the diffuse scattering in terms of transmission coefficients. Here $Q_\rp^1 \propto {T_\rp^{(0)}}(\kappa)$
and ${Q_\rp^1}' \propto {T_\rp^{(0)}}(\kappa')$.
  }
\end{figure}

To understand why the diffuse part becomes important for small distances we take a closer look at the purely evanescent components
of the diffuse Poynting vector in Eqs.~(\ref{Eq:SpecSecondOrder}) and (\ref{Eq:PoyntingDiff}). For the scattering of evanescent p-modes into evanescent p-modes this component can be written as
\begin{equation}
\begin{split}
  S^{(2)}_{\omega,\rm diff} &= - \int_{\kappa > k_0} \!\!\!\!\!\!\!\!\!\rd^2 \kappa \int_{\kappa' > k_0} \!\!\!\!\!\!\!\!\!\! \rd^2 \kappa' \,
                                 Q_\rp^1 \, g_2(|\boldsymbol{\kappa} - \boldsymbol{\kappa}'|)f_{\rp\rp}(\boldsymbol{\kappa},\boldsymbol{\kappa}';\omega)\, {Q_\rp^1}' \\
                            &\qquad + (1 \leftrightarrow 2)
\end{split}
\label{Eq:Diffdiscuss}
\end{equation}
where we have introduced the function $f_{\rp\rp}$ comprising all the factors which are not essential for the discussion of the integral and
\begin{equation}
  Q_\rp^1 = \frac{|t_{\rp}^{2}|^2}{|D_{\rp\rp}|^2}  2 \gamma \Im\bigl(r_{\rp}^{1}\bigr) \re^{-2 \gamma d}.
\end{equation}
The quantity $Q_\rp^1$ is proportional to the transmission coefficient ${T_\rp^{(0)}}$ for evanescent p-modes coming from slab $1$
which are transmitted into slab $2$ and conversely. This means that the integrand is proportional 
to $T_\rp^{(0)} (\kappa) g_2(|\boldsymbol{\kappa} - \boldsymbol{\kappa}'|) T_\rp^{(0)} (\kappa')$ describing the transmission of
a p-polarized evanescent wave with wave vector $\boldsymbol{\kappa}$ from slab $1$ to slab $2$ being scattered into a p-polarized evanescent wave with wave vector
$\boldsymbol{\kappa}'$ which is transmitted back into slab $1$. Therefore, this product is proportional to the transmission coefficient 
describing how much of the energy is
scattered back into the first slab by diffuse scattering within the second slab (see Fig.~\ref{Fig:DiffusePart}). 
In the non-retarded or quasi-static limit
the weighting factor $\gamma \exp(-2 \gamma d)$ for $Q_\rp^1$ becomes $\kappa \exp(-2 \kappa d)$, i.e., the transmission is
 the strongest if $\kappa \approx 1/d$. 
Hence, one can conclude that the integrand is large when $\kappa \approx \kappa' \approx 1/d$. It follows that 
for $|\boldsymbol{\kappa} - \boldsymbol{\kappa'}| \gg 1/d$ the
integrand is small. Since the scattering within the rough surface is approximately limited to $a_i |\boldsymbol{\kappa} - \boldsymbol{\kappa'}| \leq 1$ by the surface roughness power spectrum [see Eq.\ (\ref{Eq:RoughnessPowerSpec})],
we can conclude that for $a_i \ll d$ the diffuse component $S^{(2)}_{\omega,\rm diff}$ is comparatively small and can get large 
in the opposite limit, i.e., for $d \ll a_i$.    
In addition, due to the Fabry-P\'{e}rot denominators in $Q_\rp^1$ and  ${Q_\rp^1}'$ the diffuse contribution 
is especially large if $\kappa$ and $\kappa'$ corresponds to surface modes. Such scattering processes with $\kappa \approx \kappa'$
are reminiscent of directional scattering of surface modes~\cite{SimonGuha19976, Raether}.

\subsection{Transmission coefficient in the $(\omega,\kappa)$-plane}

To get some insight into the physical 
mechanism underlying the roughness correction, we will first discuss the transmission coefficients  ${T^{(0)}_j}$ and ${T^{(2)}_j}$ 
which are defined in Eqs.~(\ref{Eq:DefI0b}),(\ref{Eq:PoyntingDiff}),(\ref{Eq:PertResSpecprop}) and (\ref{Eq:PertResSpec}) and then we will turn to
the spectrum of the zeroth-order Poynting vector $S_\omega^{(0)}$ and the
second-order term $\langle S_{\omega}^{(2)} \rangle$. 

\begin{figure}[Hhbt]
  \epsfig{file = 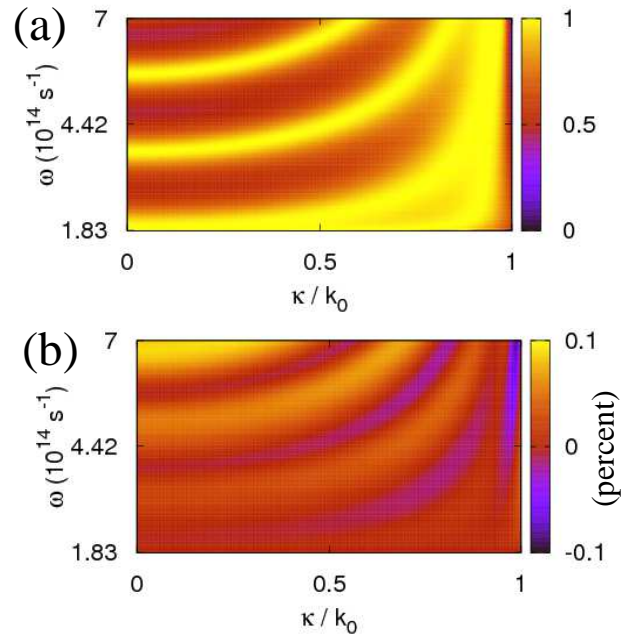, width = 0.45\textwidth}
  \caption{\label{Fig:DispersionStandingWavePrS} (Color online)
           Plot of (a) ${T^{(0)}_\rp}$ as defined in Eq.~(\ref{Eq:DefI0b}) and 
           (b) ${T^{(2)}_{\rp,\rm spec}}/{T^{(0)}_\rp}$ in the propagating regime using $d = 5\,\mu{\rm m}$ for frequencies 
           ranging from $\omega = \omega_\rl = 1.83\cdot10^{14}\,{\rm s}^{-1}$ to $7\cdot10^{14}\,{\rm s}^{-1}$. }
\end{figure}

First, for distances larger or comparable to the thermal wavelength $\lambda_{\rm th} \approx 7.68\, \mu{\rm m}$
at $300\,{\rm K}$ the propagating modes with $\kappa \leq k_0 $ dominate the
contribution to the mean Poynting vector. For the two slab configuration these modes are given by 
the gap modes for which ${T^{(0)}_j}$ equals 1 as shown in Fig.~\ref{Fig:DispersionStandingWavePrS} (a) for p-polarized modes choosing $d = 5\,\mu{\rm m}$.
The correction to the transmission coefficient due to roughness scattering is in this case dominated by the specular part. Nonetheless,
this roughness correction is rather small for the chosen roughness parameters as illustrated in Fig.~\ref{Fig:DispersionStandingWavePrS} (b).

Apart from this, due to the roughness scattering the evanescent surface polariton contribution can couple to the propagating modes~\cite{BiehsGreffet2010} so that even for
distances $d > \lambda_{\rm th}$ one finds a relatively large correction to the transmission coefficient for frequencies close to the surface resonance frequency,
which is for SiC given by $\omega_{\rm SPhP} = 1.787\cdot10^{14}\,{\rm s}^{-1}$. For $d = 5\,\mu{\rm m}$ this correction is in the order of some percent.  
Since, the propagating modes dominate the heat transfer for $d = 5\,\mu{\rm m}$ the overall correction 
to the Poynting vector is small.

\begin{figure}[Hhbt]
  \epsfig{file=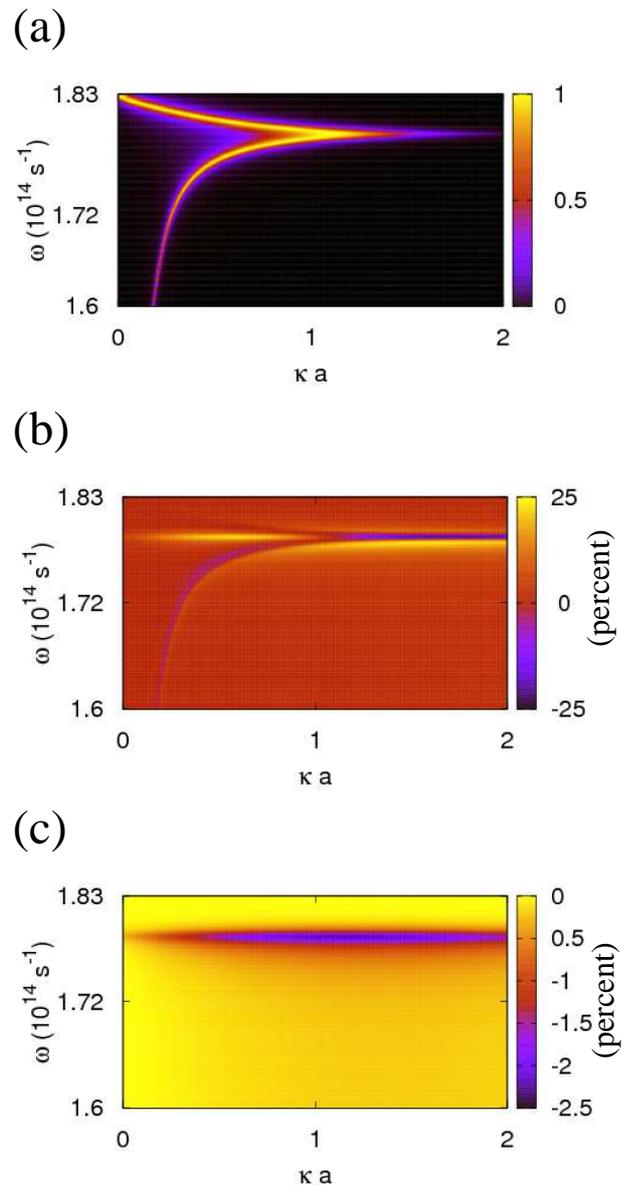, width=0.45\textwidth}
  \caption{\label{Fig:DispersionEv} (Color online)
           Plot of (a) ${T^{(0)}_\rp}$, (b) ${T^{(2)}_{\rp,\rm spec}}/{T^{(0)}_\rp}$, and
           (c) ${T^{(2)}_{\rp,{\rm diff}}}/{T^{(0)}_\rp}$ using $d = 500\,{\rm nm}$.}
\end{figure}

Now, for distances much smaller than the thermal wavelength $\lambda_{\rm th}$ the near-field heat transfer is solely 
determined by the p-polarized surface mode contribution. For such distances the diffuse contribution cannot be 
neglected. Choosing a distance of $d = 500\,{\rm nm}$ we show in 
Fig.~\ref{Fig:DispersionEv} (a) a plot of ${T^{(0)}_\rp}$ first.
Here, one can observe the splitting of the two surface phonon polariton branches~\cite{Economou1969,Raether}. 
Furthermore, one can estimate from Fig.~\ref{Fig:DispersionEv} (a) that the relevant contributions stem 
from lateral wave vectors $\kappa$
smaller than about $1.5/a$. In Fig.~\ref{Fig:DispersionEv} (b) we show the corresponding plot of the 
roughness correction ${T^{(2)}_{\rp,{\rm spec}}}/{T^{(0)}_\rp}$.  We find a qualitative similar correction as found for
the roughness scattering considering only one rough surface~\cite{BiehsGreffet2010}  
with the difference (apart from the splitting of the surface mode branches) that there is a positive correction 
for lateral 
wave vectors smaller than about $1/a$ for frequencies around $\omega_{\rm SPhP}$ and again a negative correction for lateral 
wave vectors larger than $1/a$. As we will see below, when integrating ${T^{(2)}_\rp}$ over the lateral wave vector to 
get $\langle S_\omega^{(2)}\rangle$ this
positive correction will make the negative one weaker compared to the case of only one rough surface in Ref.~\cite{BiehsGreffet2010}.
This results in a positive overall correction to the heat flux for all distances in contrast to the results found for only 
one rough surface~\cite{BiehsGreffet2010}.

In Fig.~\ref{Fig:DispersionEv} (c) we plot the corresponding diffuse contribution ${T^{(2)}_{\rp,{\rm diff}}}/{T^{(0)}_\rp}$ in the same $(\omega,\kappa)$-range.
As expected the diffuse component gives a negative correction which is relatively large for frequencies near the surface mode resonance. This
correction can be as large as $2.5$ percent for $d = 500\,{\rm nm}$. 

Hence, we can conclude that in the evanescent regime where the heat transfer is due to the coupling of surface phonon 
polaritons the roughness correction can be relatively large even for a shallow surface roughnesses.  

\subsection{Spectral roughness correction}

\begin{figure}[Hhbt]
  \epsfig{file = 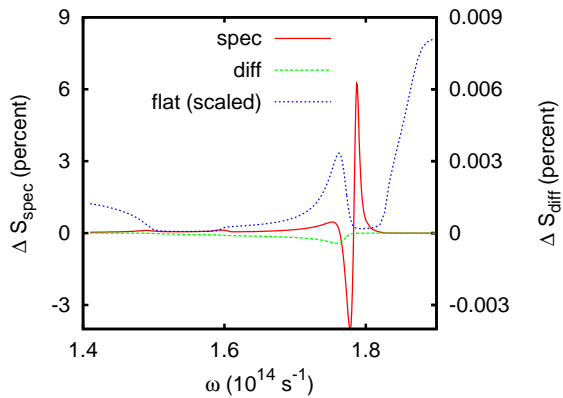 , width=0.45\textwidth}
  \caption{\label{Fig:Spectral5000nm} (Color online)
   Plot of $\langle S_\omega^{(2)} \rangle/S^{(0)}_\omega$ in a small frequency range around
   the surface phonon frequency of SiC ($\omega_{\rm SPhP} = 1.787\cdot10^{14}\,{\rm s}^{-1}$) using $d = 5\mu{\rm m}$. We have plotted the 
   specular (red solid line) and the diffuse (green dashed line) contributions (note the different scales at the right  
   and left hand side) and $S^{(0)}_\omega$ (blue dotted line).
  }
\end{figure}

Now, we turn to the effect of roughness scattering to the Poynting vector in the spectral domain, i.e., we concentrate 
on $\langle S^{(2)}_\omega \rangle / S^{(0)}_\omega$.  
In Fig.~\ref{Fig:Spectral5000nm} we have plotted $\langle S^{(2)}_\omega \rangle / S^{(0)}_\omega$ for the specular and diffuse contribution
using $d = 5\,\mu{\rm m}$ in a frequency range between $1.4\cdot10^{14}\,{\rm s}^{-1}$
and $2 \cdot 10^{14}\,{\rm s}^{-1}$. We have also plotted $S^{(0)}_\omega$ using a scaling such that the curve fits 
into the plot. First of all
one can see that, for such a large distance, the diffuse contribution is negative as implied by Eq.~(\ref{Eq:PoyntingDiff}) and 
about three orders of magnitude smaller than
the specular contribution. On the other hand, the specular contribution gives a relatively large correction of 
some percent near 
the surface resonance $\omega_{\rm SPhP} = 1.787\,{\rm s}^{-1}$, only.
This correction is the sum of the positive and negative corrections to the propagating gap mode and  
the evanescent and propagating polariton mode correction within the reststrahlen region. 
Since the positive contribution to $\langle S^{(2)}_\omega \rangle / S^{(0)}_\omega$ is larger than 
the negative one, one can expect to get a purely 
positive roughness correction to the Poynting vector 
when carrying out the frequency integral. Furthermore, since the heat flux is not solely determined by 
the surface mode, but rather dominated by the gap modes (for frequencies greater than $\omega_l$) for
which the roughness correction is very small, it follows that the roughness correction to the Poynting 
vector is small for distances around or greater than
the thermal wavelength. 

\begin{figure}[Hhbt]
  \epsfig{file =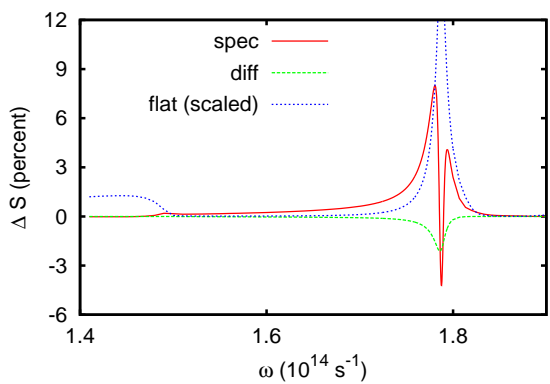 , width=0.45\textwidth}
  \caption{\label{Fig:Spectral} (Color online)
           As Fig.~\ref{Fig:Spectral5000nm} using $d = 500\,{\rm nm}$. 
  }
\end{figure}

In the evanescent regime, i.e., for a distance of $d = 500\,{\rm nm}$ we show  
in Fig.~\ref{Fig:Spectral} the correction due to the scattering of surface phonon polaritons.
As in the case for one rough surface~\cite{BiehsGreffet2010}
 we find for the specular part a positive correction for frequencies below and above the surface phonon polariton 
frequency ($\omega_{\rm SPhP} = 1.787\,{s}^{-1}$) and a negative
one for frequencies very close to the $\omega_{\rm SPhP}$. In contrast to the case of one rough surface, here 
the positive contributions are larger than the negative ones so that the overall Poynting vector  
is positive at this distance. Furthermore, the diffuse contribution to the heat flux starts to play an important
role and gives contributions in the same order of magnitude as the specular one for frequencies near $\omega_{\rm SPhP}$.
This negative correction due to the diffuse component is still so small that the integral over all frequencies remains positive.
Therefore, 
for the configuration considered here, the scattering of surface phonon polaritons does not imply a negative correction to the
heat flux as it was found for the heat transfer between a nanoparticle and a rough surface~\cite{BiehsGreffet2010}. 
In addition, since the zeroth-order $S^{(0)}_\omega$ is dominated by the surface mode contribution for distances 
much smaller than $\lambda_{\rm th}$, the roughness correction of several percent gives a change of
the heat flux of several percent. Hence, in the evanescent regime one
can expect that the roughness correction will be relatively large compared to the propagating regime. 

\subsection{Proximity approximation ($d \ll \min\{a_1,a_2\}$)}

\begin{figure}[Hhbt]
  \epsfig{file = 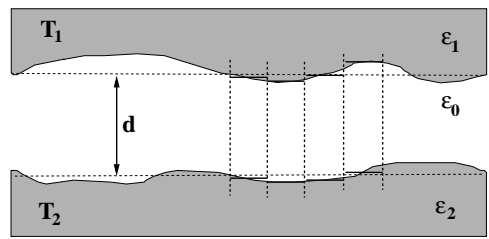, width=0.35\textwidth}
  \caption{\label{Fig:SketchPA} (Color online)
           Sketch of the proximity approximation.
  }
\end{figure}

Now, we want to compare the numerical results for the heat flux with the so-called proximity approximation (PA). 
This will be useful to understand why the coherent flux is increased by the roughness.
Within this approximation the rough surfaces are locally approximated by a flat surface (see Fig.~\ref{Fig:SketchPA}) so that the mean flux can be computed
from the flux between two flat surfaces $S^{(0)}$ by 
\begin{equation}
\begin{split}
  \langle S(d) \rangle &\approx \langle S^{(0)}(d - S_1(\mathbf{x}) + S_2(\mathbf{x})) \rangle \\
                           &=  S^{(0)}(d) + \frac{1}{2}\frac{\partial^2}{\partial d^2} S^{(0)} (d) (\delta_1^2 + \delta_2^2) + \mathcal{O}(4) \\
                           &= S_{\rm PA}.
\end{split}
\label{Eq:PAges}
\end{equation}
When considering dielectric materials, then $S^{(0)} \propto d^{-2}$ in the quasi-static regime so that the PA can be further
simplified to
\begin{equation}
  S_{\rm PA} \approx S^{(0)}\biggl( 1 + 3 \frac{(\delta_1^2 + \delta_2^2)}{d^2} \biggr). 
\label{Eq:PA}
\end{equation} 
For the case of two rough surfaces the PA was first used to estimate the near-field radiative heat transfer between two 
rough surfaces in Ref.~\cite{PerssonEtAl2010}. 

\begin{figure}[Hhbt]
  \epsfig{file = 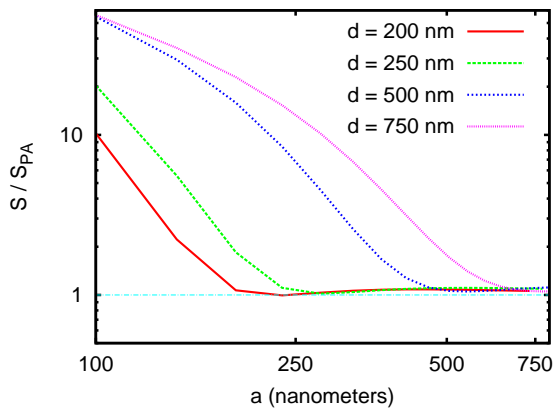 , width=0.45\textwidth}
  \caption{\label{Fig:ValidityPA} (Color online)
           Plot of $S(d)/S_{\rm PA} (d)$ for SiC over $a$ keeping the rms constant at $\delta = 5\,{\rm nm}$ for $d = 200\,{\rm nm}, 250\,{\rm nm}, 500\,{\rm nm}$, 
           and $750\,{\rm nm}$.
  }
\end{figure}

Now, with the perturbation result we can discuss the range of validity
of this very simple approximation. As for the case of one rough surface discussed in Ref.~\cite{BiehsGreffet2010} one can expect that
the PA is valid as far as $d \ll \min\{a_1, a_2\}$, since then the local approximation by flat surfaces with a lateral extension much smaller than $a_i$ can be assumed to be useful.
In order to explore the range of validity of the PA we show in Fig.~\ref{Fig:ValidityPA} a plot of $S(d)/S_{\rm PA} (d)$ over the correlation length $a_1 = a_2 = a$ for different distances. It can be
seen that the ratio $S(d)/S_{\rm PA} (d)$ goes approximately to $1$ for $a > d$. Hence, we can conclude that the PA can be used for $d < a$.

\begin{figure}[Hhbt]
  \epsfig{file= 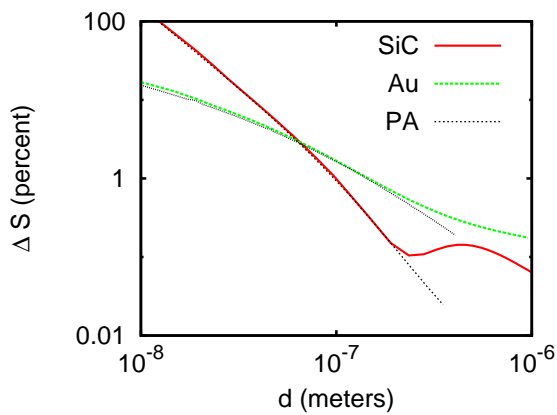 , width=0.45\textwidth}
  \caption{\label{Fig:PA} (Color online)
           As Fig.~\ref{Fig:P0} (b) but with the PA for SiC and Au calculated from $S^{(0)}$ with Eq.~(\ref{Eq:PAges}).  
  }
\end{figure}

This conclusion is not trivial, since for SiC the near-field heat transfer is due to the coupling of thermally excited surface modes. The
 thermal near fields associated with these surface modes can have a lateral coherence length $l_{\coh}$ larger than the vacuum 
wave length~\cite{CarminatiGreffet1999,JoulainEtAl2005} and can therefore be larger than the correlation length $a$ considered here.
However, in the quasi-static regime ($d \ll \lambda_{\rm th}$) it is found~\cite{HenkelEtAl2000,LauEtAl2007} 
that $l_{\coh} \approx d$. Hence, for distances $d \ll a$ the thermally generated fields above different surface elements having an area smaller than $a \times a$ and
larger than $d \times d$ are uncorrelated. Hence, for $d \ll a$ the flat areas can indeed be regarded as independent so that the PA is valid. 
 
Numerically, we find for SiC that for distances $d \ll a$ the correction to the heat flux is 
given by the sum of the positive specular and the negative diffuse contribution. The specular correction tends to give results larger than
the one predicted by the PA. Only, when also considering the diffuse component of the heat flux, we retrieve the PA result. 
Therefore, the interplay of the diffuse and the specular scattering of the surface polaritons guarantees the validity of the PA. 
On the other hand, for Au which has no surface resonance around $\lambda_{\rm th}$ we find that the diffuse part is negligible small. 
In this case the PA is also valid and is due to the specular part only. 

In Fig.~\ref{Fig:PA} we plot the roughness correction for SiC and Au together with 
a numerical evaluation of the PA in Eq.~(\ref{Eq:PAges}). In both cases the PA gives a good approximation  
for distances $d \ll a$. As can be expected the curve for SiC converges for very small distances to the 
quasi-static expression for the PA in Eq.~(\ref{Eq:PA}), since $S^{(0)} \propto 1/d^2$ in the quasi-static regime. For
Au $S^{(0)}$ saturates for small distances so that the quasi-static expression in Eq.~(\ref{Eq:PA}) is not valid in that case. 

To summarize, for distances $d \ll \min\{a_1,a_2\}$ the PA gives a good approximation of the impact of surface roughness for all the materials considered here. This
allows for estimating the effect of surface roughness from the results obtained for a flat surface and is therefore an important result of this work. Note,
that the same range of validity for the PA has also been reported for the Casimir force~\cite{GenetEtAl2003, MaiaNetoEtAlLett2005, MaiaNetoEtAl2005}.
In this case, it has also been demonstrated that the theoretical predictions using the PA are in good
agreement with the experimental data~\cite{ChanEtAl2001,KlimtchitskayaEtAl2009}.

%
%

\section{Conclusion}

We have presented a perturbation theory for determining the surface roughness
correction to the heat flux between two semi-infinite bodies. In particular, we have given the
explicit expression for the mean Poynting vector up to second-order perturbation theory assuming
a Gaussian surface roughness. These results can be employed to estimate the influence of roughness in 
recent experiments~\cite{HuEtAl2008,NarayaEtAl2008,ShenEtAl2008,RousseauEtAl2009} and to study the impact of
roughness in thermophotovoltaic devices~\cite{MatteoEtAl2001, NarayanaswamyChen2003, LarocheEtAl2006, FrancoeurEtAl2008,ZhangReview}.

In addition, we have discussed the numerical results for two SiC slabs with a given surface roughness.
We found that the heat flux becomes larger when surface roughness is taken into account.
By giving  a detailed discussion of the correction to the transmission coefficients and the spectral Poynting
vector for two SiC slabs, we could show that the scattering of surface modes within the rough surfaces causes a larger heat flux
which is in contrast to the case of only one rough surface as discussed in Ref.~\cite{BiehsGreffet2010}. 

Finally, we have shown that the PA is valid for distances smaller than the correlation length of the surface roughness regardless of the
materials (SiC, Au) considered in this work. In particular, for the case of two SiC slabs the validity of the PA is due to an interplay
of the specular and diffuse contribution to the mean Poynting vector, whereas for Au which do not have a surface polariton resonance
in the infrared region 
the specular part already gives the PA result. Hence, the PA is valid even when the heat flux is mainly due to the surface polariton coupling giving
a simple and therefore powerful way of estimating the impact of surface roughness to the heat flux.

%
%

\begin{acknowledgments}
S.-A.\ B. gratefully acknowledges support from the Deutsche Akademie der Naturforscher Leopoldina
(Grant No.\ LPDS 2009-7).
\end{acknowledgments}

%
%

\appendix

%
%

\end{document}